\documentclass[12pt]{iopart}

\usepackage{graphicx}
\usepackage{dcolumn}
\usepackage{bm}
\usepackage{hyperref}

\usepackage{multirow}%
\usepackage{amsmath,amssymb,amsfonts}%
\usepackage{tabularx,booktabs}
\usepackage{mathtools}
\usepackage{amsthm}

\newtheorem*{theorem}{Theorem}

\newcommand{\RNum}[1]{\uppercase\expandafter{\romannumeral #1\relax}}

\begin{document}

\title[Warped Product Einstein Manifolds in Four Dimensions]{Warped Product Einstein Manifolds in Four Dimensions}

\author{Jack C. M. Hughes$^{1,2}$, Joudy F. Jamal Beek$^{2}$ \& Fedor V. Kusmartsev$^{2}$}

\address{
  \small $^{1}$Department of Public Health and Epidemiology, College of Medicine and Health Sciences, Khalifa University, PO Box 127788, Abu Dhabi, United Arab Emirates\\
  \small $^{2}$Department of Physics, College of Engineering and Physical Sciences, Khalifa University, PO Box 127788, Abu Dhabi, United Arab Emirates}
\ead{jack.hughes@ku.ac.ae}
\vspace{10pt}
\begin{indented}
\item[]May 2026
\end{indented}

\begin{abstract}
On four-dimensional (pseudo-)Riemannian manifolds $\mathcal{M}$ the curvature tensor (viewed as an endomorphism on 2-forms) admits a chiral $6 \times 6$ matrix representation which decomposes into four $3 \times 3$ blocks. $\mathcal{M}$ is Einstein if and only if the off-diagonal blocks vanish. If the manifold is a warped product $\mathcal{M} = F \times_f B$, then there exists an alternative matrix representation relative to the decomposition of the 2-forms into spaces induced by the exterior algebra on both the base and the fiber. These two representations are not independent and a similarity transformation can be found between them. We construct these matrices and associated transformations for $1+3$, $2+2$ and $3+1$ warped products, giving classifications for the Einstein limits from this algebraic perspective. Using this, one can easily Petrov classify the Einstein warped products for each case considered: $3+1$ are generically type-\RNum{1}, $2+2$ are type-D while $1+3$ are constrained to be type-O. In the closed Riemannian case, there are a number of topological restrictions on these manifolds that we discuss: in the half-conformally flat limit, each of these Einstein warped products must be flat.
\end{abstract}

\section{Introduction}
In this work we will be interested in matrix representations of the curvature tensor on four-dimensional warped products. There are a number of reasons why this deserves attention. The first is that many of the `classical' spacetimes of note are of course warped products \cite{straumann_general_2013, fre_gravity_2013, heller_theoretical_1992}: the FLRW cosmologies, spherically symmetric spacetimes, near-horizon geometries in Euclidean signature (especially in the context of the Gibbons-Hawking entropy calculation via Wick rotation) \cite{gibbons_action_1977, diakonov_sitter_2025, volovik_effect_2021, hughes_topological_2025, cai_action_1998}, topological black holes \cite{cai_gauss-bonnet_2002, cai_note_2004, birmingham_topological_1999, anninos_warped_2009, vanzo_black_1997}, Randall-Sundrum models \cite{randall_large_1999, randall_alternative_1999}, etc. The second reason is that spacetimes in four dimensions admit a number of special properties by virtue of the fact that the Lie algebra of the Lorentz group has a semi-simple decomposition \cite{atiyah_self-duality_1978, besse_einstein_1987, woit_euclidean_2021, krasnov_formulations_2020}
\begin{equation}\label{Lie_Algebra_Decomp}
    \mathfrak{so}_{\mathbb{C}}(1,3) \cong \mathfrak{sl}(2,\mathbb{C}) \oplus \overline{\mathfrak{sl}(2,\mathbb{C})}, \quad \mathfrak{so}(4) \cong \mathfrak{su}(2) \oplus \mathfrak{su}(2).
\end{equation}
In particular, the space of 2-forms decomposes into two orthogonal 3-dimensional subspaces
\begin{equation}
    \Lambda^2 = \Lambda^+ \oplus \Lambda^-
\end{equation}
characterized by the eigenvalues of the Hodge dual (complex in spacetime signature). In four dimensions only, one may characterize the `vacuum' state of general relativity (GR) by the fact that the curvature of the self-dual $\Lambda^+$ part of the connection is \textit{purely self-dual} as a 2-form \cite{atiyah_self-duality_1978, capovilla_self-dual_1991}. This is what allows for the construction of the Plebanski and chiral pure connection formalisms of GR \cite{krasnov_formulations_2020, plebanski_separation_1977, shaw_general_2026, krasnov_pure_2011}, together with the associated interest in the heavenly equations, spinors and twistors \cite{woit_euclidean_2021, krasnov_pure_2025, alexandrov_self-dual_2010, plebanski_solutions_1975, manakov_inverse_2006, mason_integrability_1996, bittleston_twistors_2023}. (\ref{Lie_Algebra_Decomp}) is also the basis for the Petrov classification of the Weyl tensor (viewed as an endomorphism on the self-dual bivectors) \cite{petrov_classification_2000, stephani_exact_2003}.

The relation between the chiral curvature decomposition and warped-product curvature blocks appears to have received comparatively little direct attention. While warped product Einstein manifolds have been of interest \cite{de_sousa_family_2017, aloui_einstein_2025, he_classification_2011, leandro_structure_2018, he_uniqueness_2013, kim_warped_2006}, with several impressive results being obtained \cite{ferrando_intrinsic_2010, carot_geometry_1993, chen_survey_2013, an_warped_2017, du_surfaces_2021, minguzzi_causal_2007, ortaggio_higher_2011, allison_geodesic_1988, unal_doubly_2001}, it is typically with an eye for generality in mind, with four dimensions no more special than any other choice. The perspective adopted here is that the warped-product splitting and the chiral splitting of 2-forms provide two alternative but linearly related decompositions of the algebraic curvature operator. The former is adapted to the metric product structure, while the latter is adapted to the Hodge duality underlying the Einstein condition and the Plebanski formulation. Passing between these two descriptions by explicit similarity transformations allows the Einstein condition to be easily obtained as a concrete algebraic constraint on the warped product curvature blocks. This gives a compact approach to seeing why the three possible four-dimensional splittings $(1+3), (2+2)$ and $(3+1)$ behave in fundamentally different ways (namely via their interaction with duality), despite all being warped products of total dimension four.  
The resulting algebraic classification \cite{petrov_classification_2000} is rigid in the (1+3) case, where the Einstein condition forces constant curvature; it is intermediate in the (2+2) case, where the Einstein limit gives a type-D chiral structure; and is most flexible in the (3+1) case, where the Einstein condition equates two $(3\times 3)$ curvature blocks without forcing them to be scalar multiples of the identity.

\section{Preliminaries}
Before proceeding, it is worth recalling the essential results regarding four dimensional manifolds and (separately) warped products. This will allow for a fixing of conventions. In what follows, we will work for simplicity in Riemannian signature. The spacetime case proceeds extremely similarly and involves only additional minus signs under the Hodge dual and complexification of the matrices in the chiral decompositions. The conclusions for the forms of the Einstein warped product matrices are the same regardless of signature.

\subsection{The Riemann Matrix}
In GR we are typically interested in manifolds $\mathcal{M}$ equipped with a metric $g$ and a torsion free, metric compatible connection $\nabla$ \cite{krasnov_formulations_2020, fre_gravity_2013-1}. With these restrictions, the Riemann tensor $R_{\mu \nu \rho \sigma}$ possesses a number of important symmetries:
\begin{equation}\label{Symmetries_Rie}
    R_{\mu \nu \rho \sigma} = -R_{\nu \mu \rho \sigma} = - R_{\mu \nu \sigma \rho} = R_{\rho \sigma \mu \nu}.
\end{equation}
One way to understand this is by raising a pair of antisymmetric indices and consequently viewing the Riemann tensor as a map from the space of 2-forms $\Lambda^2$ into itself (hence an endomorphism),
\begin{equation}\label{Riemann_Map}
    \begin{split}
        R_{\mu \nu}^{\quad \rho \sigma} \colon \Lambda^2 & \to \Lambda^2 \\
        \omega_{\rho \sigma}  &\mapsto \omega'_{\mu \nu} = R_{\mu \nu}^{\quad \rho \sigma}\omega_{\rho \sigma}.
    \end{split}
\end{equation}
However, since 
\begin{equation}
    R_{\mu \nu}^{\quad \rho \sigma} = R^{\rho \sigma}_{\quad \mu \nu}
\end{equation}
we note that the mapping (\ref{Riemann_Map}) constitutes a $\dim(\Lambda^2) \times \dim(\Lambda^2)$ symmetric matrix. 

As an example, consider three dimensions. The space of 2-forms $\Lambda^2$ is three-dimensional. Since the Weyl tensor vanishes identically, the Riemann tensor is completely determined by the Ricci tensor:
\begin{equation}\label{3d_Decomp}
    R_{\mu \nu}^{\quad \rho \sigma} = 2\big(\delta_{\mu}^{[\rho}R_{\nu}^{\sigma]} - \delta_{\nu}^{[\rho} R_{\mu}^{\sigma]} \big) - R\delta_{\mu}^{[\rho} \delta_{\nu}^{\sigma]}.
\end{equation}
In the standard orthonormal co-frame $(e^1, e^2, e^3)$, which spans $\Lambda^2$ via the wedge products 
\begin{equation}
    \Omega_1 = e^2 \wedge e^3, \quad \Omega_2 = e^3 \wedge e^1, \quad \Omega_3 = e^1 \wedge e^2,
\end{equation}
it is not difficult to see (with $i,j$ labeling the $\Omega_i$ basis) that as a matrix we have 
\begin{equation}\label{3d_Matrix}
    \text{Riem}_{ij} = \mathcal{R}_{ij} - \frac{R}{2}\mathbb{I}_{ij},
\end{equation}
where
\begin{equation}\label{3d_Curv_Operators}
     \mathbb{I}_{ij} = \big(\delta_{\mu}^{[\rho} \delta_{\nu}^{\sigma]}\big)_{ij}, \quad \mathcal{R}_{ij} = 2\big(\delta_{\mu}^{[\rho}R_{\nu}^{\sigma]} - \delta_{\nu}^{[\rho} R_{\mu}^{\sigma]} \big)_{ij}.
\end{equation}
For instance, $\text{Riem}_{11}$ is a map 
\begin{equation}
    \text{Riem}_{11} \colon \Omega_1 \to \Omega_1.
\end{equation}
At the tensor level (\ref{3d_Decomp}), this is generated via
\begin{equation}
    R_{23}^{\quad 23} = 2\big(\delta_{2}^{[2}R_{3}^{3]} - \delta_{3}^{[2} R_{2}^{3]} \big) - R\delta_{2}^{[2} \delta_{3}^{3]} = \delta^2_2 R^3_3 + \delta^3_3 R^2_2 - \frac{R}{2} \delta^2_2 \delta^3_3.
\end{equation}

Notice that the matrix representation allows for an easy characterization of what it means for $\mathcal{M}$ to be Einstein (and hence satisfy the vacuum field equations). Indeed, a manifold is said to be Einstein if \cite{besse_einstein_1987}
\begin{equation}\label{Einstein_Cond}
    R_{\mu \nu} = \Lambda g_{\mu \nu}
\end{equation}
for some constant $\Lambda \in \mathbb{R}$. This immediately implies that the matrix (\ref{3d_Matrix}) is diagonal,
\begin{equation}\label{Riem_3d_Einstein}
    \text{Riem}_{ij} = \frac{\Lambda}{2} \mathbb{I}_{ij}.
\end{equation}
Obviously this case is quite simple, since in three dimensions Einstein manifolds are constant curvature. It is in fact easy to show from the above that a 3-manifold is Einstein if and only if the Riemann matrix is a constant multiple of the identity\footnote{More generally of course, a manifold (in any dimension) is constant curvature if and only if the Riemann matrix is a constant multiple of the identity.}. The various ways one can enforce this condition variationally provide the origin of the topological Chern-Simons formulation \cite{krasnov_formulations_2020, wise_symmetric_2009, mielke_geometrodynamics_2017}.

Transitioning to four dimensions, the $6\times 6$ matrix form of the Riemann operator (\ref{Riemann_Map}) is determined similarly by its orthogonal decomposition
\begin{equation}\label{Irreducible_Decomp_Riem}
    \text{Riem}_{ij} = \mathcal{W}_{ij} + Z_{ij} + \frac{R}{12} \mathbb{I}_{ij},
\end{equation}
where $(i,j)$ now labels the standard $\Lambda^2$ basis spanned by
\begin{equation}\label{4d_Basis}
    \begin{split}
        \Omega_1 = e^1 \wedge e^2, \quad \Omega_2 = e^1 \wedge e^3, \quad \Omega_3 = e^1 \wedge e^4, \\
        \Omega_4 = e^3 \wedge e^4, \quad \Omega_5 = e^4 \wedge e^2, \quad \Omega_6 = e^2 \wedge e^3,
    \end{split}
\end{equation}
giving the matrix representation of the $\Lambda^2$ endomorphism operators 
\begin{equation}\label{Curv_Operators}
    \mathcal{W}_{ij} = \big(C_{\mu \nu}^{\quad \rho \sigma}\big)_{ij}, \quad Z_{ij} = \frac{1}{2} \big(\delta_{\mu}^{[\rho}Z^{\sigma]}_{\nu} - \delta_{\nu}^{[\rho} Z_{\mu}^{\sigma]})_{ij}, \quad \mathbb{I}_{ij} = \big(\delta^\rho_{[\mu} \delta^{\sigma}_{\nu]}\big)_{ij},
\end{equation}
where $C_{\mu \nu \rho \sigma}$ is the Weyl tensor and $Z_{\mu \nu}$ is the trace-free Ricci tensor \cite{besse_einstein_1987}. At this point, the matrix $\text{Riem}_{ij}$ is complicated and has in general no vanishing elements (thanks to the non-trivial structure of the Weyl tensor). When $\mathcal{M}$ is Einstein (\ref{Einstein_Cond}), then the trace-free Ricci tensor vanishes, hence $Z_{ij} = 0$. Again, this condition is if and only if \cite{besse_einstein_1987, krasnov_formulations_2020}. However, unlike for the three-dimensional case (\ref{Riem_3d_Einstein}) there is not a simple condition on the matrix structure of $\text{Riem}_{ij}$ within the basis (\ref{4d_Basis}). Fortunately, the presence of \textit{duality} within $\Lambda^2$ allows for much more to be said.

\subsection{Duality in Four Dimensions}

In any dimension \textit{other than four}, the space $\Lambda^2$ is irreducible under the action of the special orthogonal group \cite{besse_einstein_1987, hughes_encoding_2026}. However, in four dimensions the Hodge dual also acts as an endomorphism
\begin{equation}\label{Hodge_Map}
    \begin{split}
        \star \colon \Lambda^2 & \to \Lambda^2 \\
        \omega_{\rho \sigma}  &\mapsto (\star\omega)_{\mu \nu} = \frac{1}{2} \epsilon_{\mu \nu}^{\quad \rho \sigma}\omega_{\rho \sigma}
    \end{split}
\end{equation}
with the special property that it squares to (a multiple of) the identity,
\begin{equation}\label{Hodge_Square}
    (\star^2 \omega)_{\mu \nu} = \frac{1}{4} \epsilon_{\mu \nu}^{\quad \rho \sigma} \epsilon_{\rho \sigma}^{\quad \alpha\beta} \omega_{\alpha \beta} = \pm \delta_{[\mu}^{\alpha} \delta_{\nu]}^{\beta} \omega_{\alpha \beta} = \pm\omega_{\mu\nu},
\end{equation}
where the sign here depends on the signature (negative for spacetime) \cite{krasnov_formulations_2020}. As such, we can further decompose $\Lambda^2$ into the associated orthogonal eigenspaces of the Hodge dual,
\begin{equation}\label{2_Form_Decomp}
    \Lambda^2 = \Lambda^+ \oplus \Lambda^-.
\end{equation}
Each of these subspaces are three-dimensional, which will allow us to construct a matrix representation
\begin{equation}\label{Chiral_Matrix}
    \text{Riem}' = \begin{pmatrix}
        \mathcal{A} & \mathcal{B} \\
        \mathcal{B}^T & \mathcal{C}
    \end{pmatrix}
\end{equation}
into $3 \times 3$ blocks adapted to these subspaces. Since the basis elements in (\ref{4d_Basis}) are dual to eachother (e.g. $\star \Omega_1 = \Omega_4$), a convenient choice of basis for the decomposition (\ref{2_Form_Decomp}) is (in Euclidean signature)
\begin{equation}\label{Chiral_Basis}
    \begin{split}
        \Sigma_1^+ = \frac{1}{\sqrt{2}}( \Omega_1 + \Omega_4), \quad \Sigma_2^+ = \frac{1}{\sqrt{2}}( \Omega_2 + \Omega_5), \quad \Sigma_3^+ = \frac{1}{\sqrt{2}}( \Omega_3 + \Omega_6), \\
        \Sigma_1^- = \frac{1}{\sqrt{2}}( \Omega_1 - \Omega_4), \quad \Sigma_2^- = \frac{1}{\sqrt{2}}( \Omega_2 - \Omega_5), \quad \Sigma_3^- = \frac{1}{\sqrt{2}}( \Omega_3 - \Omega_6).
    \end{split}
\end{equation}
The transformation matrix mapping the basis (\ref{4d_Basis}) into the \textit{chiral} basis (\ref{Chiral_Basis}) is
\begin{equation}\label{P_Matrix}
    \mathcal{P} = \frac{1}{\sqrt{2}}\begin{pmatrix}
        1 & 0 & 0 & 1 & 0 & 0 \\
        0 & 1 & 0 & 0 & 1 & 0 \\
        0 & 0 & 1 & 0 & 0 & 1 \\
        1 & 0 & 0 & -1 & 0 & 0 \\
        0 & 1 & 0 & 0 & -1 & 0 \\
        0 & 0 & 1 & 0 & 0 & -1
    \end{pmatrix}
\end{equation}
Note that since $\mathcal{P}$ is orthogonal, $\mathcal{P}^{-1} = \mathcal{P}^T$. This allows us to deduce the geometric content of the blocks of $\text{Riem}'_{ij}$ via a similarity transformation 
\begin{equation}
    \text{Riem}' = \mathcal{P} \; \text{Riem} \; \mathcal{P}^T.
\end{equation}
A sequence of facts now becomes essential for this purpose:
\begin{enumerate}
    \item In the chiral basis (\ref{Chiral_Basis}), the Hodge dual operator (\ref{Hodge_Map}) is \textit{diagonal},
    \begin{equation}\label{Hodge_Matrix}
        \star = \begin{pmatrix}
            \mathbb{I}_3 & 0 \\
            0 & -\mathbb{I}_3
        \end{pmatrix},
    \end{equation}
    where $\mathbb{I}_3$ is the identity operator (\ref{Hodge_Square}) restricted to the three-dimensional subspaces $\Lambda^+, \Lambda^-$.
    \item The Weyl tensor (viewed as an endomorphism on 2-forms, (\ref{Curv_Operators})) \textit{commutes with the Hodge dual},
    \begin{equation}
        C_{\mu \nu}^{\quad \rho \sigma} \epsilon_{\rho \sigma}^{\quad \alpha \beta} = \epsilon_{\mu \nu}^{\quad \rho \sigma} C_{\rho \sigma}^{\quad \alpha \beta}.
    \end{equation}
    This follows at the tensorial level from the symmetries of the Weyl tensor (\ref{Symmetries_Rie}) together with the fact that it is traceless \cite{kroon_conformal_2023}. At the matrix level, (\ref{Hodge_Matrix}) then implies 
    \begin{equation}
        \mathcal{W}' = \mathcal{P} \; \text{Weyl} \; \mathcal{P}^T, \quad [\star, \mathcal{W}'] = 0
    \end{equation}
    or that $\mathcal{W}$ is block diagonal in the chiral basis. Let us call these blocks respectively
    \begin{equation}\label{Weyl_Matrix}
        \mathcal{W}' = \begin{pmatrix}
            C^+ & 0 \\
            0 & C^-
        \end{pmatrix}.
    \end{equation}
    \item The tensor
    \begin{equation}
        \mathcal{Z}_{\mu \nu}^{\quad \rho \sigma} = \delta_{\mu}^{[\rho}Z^{\sigma]}_{\nu} - \delta_{\nu}^{[\rho} Z_{\mu}^{\sigma]}
    \end{equation}
    \textit{anticommutes with the Hodge dual},
    \begin{equation}
        \mathcal{Z}_{\mu \nu}^{\quad \rho \sigma} \epsilon_{\rho \sigma}^{\quad \alpha \beta} = -\epsilon_{\mu \nu}^{\quad \rho \sigma} \mathcal{Z}_{\rho \sigma}^{\quad \alpha \beta}.
    \end{equation}
    This follows simply from a standard four-dimensional $\epsilon$ identity. Since this tensor is part of the endomorphism operators (\ref{Curv_Operators}) forming the Riemann matrix (note that it is simply the Kulkarni-Nomizu lift of the trace-free Ricci tensor \cite{hughes_encoding_2026}), we have
    \begin{equation}
        \mathcal{Z}' = \mathcal{P} \; \mathcal{Z} \; \mathcal{P}^T, \quad \{\star, \mathcal{Z}'\} = 0,
    \end{equation}
    or that $\mathcal{Z}'$ is off-block diagonal in the chiral basis.
\end{enumerate}
Taking these facts together, we have that in the chiral basis 
\begin{equation}\label{Chiral_Matrix_End}
    \text{Riem}' = \begin{pmatrix}
        C^+ + \frac{R}{12} \mathbb{I}_3 & \mathcal{Z}' \\ \mathcal{Z}'^T & C^- + \frac{R}{12} \mathbb{I}_3
    \end{pmatrix}
\end{equation}
The point now is that within this basis, there is a simple presentation of the Einstein condition: a manifold is Einstein \textit{if and only if} the matrix representation of the Riemann tensor in the chiral basis is block diagonal, $\mathcal{Z}' = 0$ \cite{besse_einstein_1987, krasnov_formulations_2020, hughes_encoding_2026}. We will use similar matrix algebra below when relating warp product decompositions of $\Lambda^2$ to the chiral basis (\ref{Chiral_Matrix_End}).

\subsection{Warped Products}
A warped product manifold is a product manifold $\mathcal{M} = B \times_f F$ where the metric $g$ on $\mathcal{M}$ is of the form
\begin{equation}\label{Warp_Metric}
    g = g_B + f^2 g_F,
\end{equation}
where $f\colon B \to \mathbb{R}^+$ is a smooth positive function (the warping function) of the coordinates on $B$ \cite{allison_lorentzian_1985, oneill_semi-riemannian_1983, beem_global_1981}. If $\dim B = p$ and $\dim F = q$, then $\dim \mathcal{M} = p + q$. Using this, one can derive all the standard expressions for the curvature tensors of interest on $\mathcal{M}$ in terms of the warping function $f$. In adapted coordinates ($x^a, y^i$), the curvature tensor on $\mathcal{M}$ decomposes into the curvature of $B$ \cite{oneill_semi-riemannian_1983}
\begin{equation}\label{Warp_R_1}
    R_{abcd} = (R^B)_{abcd},
\end{equation}
the curvature due to mixing
\begin{equation}\label{Warp_R_2}
    R_{aibj} = - \frac{\nabla_a \nabla_b f}{f} g_{ij}
\end{equation}
and the curvature on $F$ together with the warping factor
\begin{equation}\label{Warp_R_3}
    R_{ijkl} = f^2 (R^F)_{ijkl} - \frac{\vert \nabla f \vert^2}{f^2} (g_{ik} g_{jl} - g_{il} g_{jk})
\end{equation}
(with $R_{abci}=0$ and $R_{aijk} = 0$). At this point one can begin to rephrase various constraints on the structure of $\mathcal{M}$ in terms of the curvature decomposition relative to the warp (e.g. maximally symmetric, parallel and so on). In the case where $\mathcal{M} = B \times_f F$ is restricted to be Einstein (\ref{Einstein_Cond}), one gets the following constraints on the Ricci tensors on the factors:
\begin{gather}
    (\text{Ric}^B)_{ab} - \Lambda (g_B)_{ab} = n \frac{\nabla_a \nabla_b f}{f}, \\
    (\text{Ric}^F)_{ij} = \lambda_F (g_F)_{ij}, \\
    \lambda_F = \Lambda f^2 + \frac{f^2}{n} (R_B -m\Lambda) + (n-1) \vert \nabla f\vert^2.
\end{gather}
In words, the first equation states that the failure of the base $B$ to be Einstein is quantified by the (normalized) Hessian in $f$. The second equation states that the fiber $F$ is Einstein, with the third equation giving the constant $\lambda_F$ in terms $\Lambda$, the warping function $f$ and the Ricci scalar $R_B$ on the base.

For our purposes, it is interesting to recognize that the tangent bundle splits orthogonally as 
\begin{equation}
    T\mathcal{M} = TB \oplus TF,
\end{equation}
which immediately induces a decomposition on the exterior algebra. In particular, for the 2-forms
\begin{equation}\label{2_Form_Split_Warp}
    \Lambda^2 = \Lambda^2 T^*B \oplus \Lambda^2 T^* F \oplus \big( T^*B \wedge T^* F\big)
\end{equation}
Symmetrizing over this space, we have
\begin{equation}\label{Warp_Alg_Curv}
    \begin{split}
        S^2(\Lambda^2 T^*\mathcal{M}) & \cong S^2(\Lambda^2 T^* B) \oplus S^2 (\Lambda^2 T^* F) \oplus S^2\big( T^*B \wedge T^* F\big) \oplus \\
        & (\Lambda^2 T^*B \otimes (T^*B \wedge T^*F)) \oplus (\Lambda^2 T^* B \otimes \Lambda^2 T^* F) \oplus ((T^*B \wedge T^*F) \otimes \Lambda^2 T^*F).
    \end{split}
\end{equation}
The first Bianchi identity, $R_{[abcd]} = 0$, places a restriction on the completely antisymmetric subspace within $S^2 (T^*B \wedge T^*F)$ and $\Lambda^2 T^* B \otimes \Lambda^2 T^*F$ (we will return to this once we specify dimensions). \textit{The algebraic space of curvature tensors} $\mathcal{K}$ (the space containing the Riemann curvature) is the space $S^2(\Lambda^2 T^*\mathcal{M})$ equipped with this Bianchi constraint \cite{besse_einstein_1987, hughes_encoding_2026}. Relative to the warp then, this gives a decomposition of the Riemann matrix (\ref{Riemann_Map}) that is distinct (but not independent) to the standard orthogonal (\ref{Curv_Operators}) and chiral (\ref{Chiral_Matrix}) decompositions. The exact form of the matrix depends on the dimensions of $B$ and $F$. However, due to the unique structure of the curvature for the warp (\ref{Warp_R_1})-(\ref{Warp_R_3}), many of the irreducible representations in (\ref{Warp_Alg_Curv}) collapse and the matrix structure is particularly transparent. It is worth pointing out (though not particularly useful here) that certain elements of this irreducible decomposition can be constructed via the Kulkarni-Nomizu product between symmetric tensors on the base and fiber, together with the both the base and fiber metrics. 

Below we will construct this matrix for the three cases of interest in four dimensions, using bases of $\Lambda^2$ adapted to the warp. We will then develop the similarity transformations to the chiral basis (\ref{Chiral_Basis}) and interpret the Einstein condition (namely that the Riemann matrix in the chiral basis commutes with the Hodge dual) as a constraint on the structure of the warp. One can then provide structural theorems on the eigenvalues of the matrix in the warped basis for the Einstein condition of the `if and only if' form.

\section{Four-dimensional Warped Products}
There are three cases of interest for a four-dimensional warp, and each yield comparatively different dimensional decompositions of the space of algebraic space of curvature tensors. In both the 1 + 3 case (relevant for, say, cosmology) and the $3+1$ case (static manifolds) one gets a $\mathbf{3 \oplus 3}$ splitting of $\Lambda^2$, very similar to the chiral decomposition (\ref{2_Form_Decomp}). Despite this fact, these two possibilities prove to be quite different. However, for the $2+2$ case we instead have $\mathbf{1 \oplus 1 \oplus 4}$, which is again distinct. The structure of the Riemann matrix is in all cases unique, with the associated Einstein condition yielding alternative forms for the map (\ref{Riemann_Map}). 

\subsection{Case \RNum{1}: 1 + 3 Warp}
When the warped product contains a one-dimensional base $B$ with a three-dimensional fiber $F$, the space of 2-forms on $\mathcal{M}$ contains no `pure' elements from $B$. Instead, there is a `mixed' subspace composed of wedge products of the single 1-form on $B$ and the three 1-forms on $F$, together with the three `pure' 2-forms on the fiber. Thus,
\begin{equation}\label{1_plus_3_2_Forms_Split}
    \Lambda^2 = \Lambda^2_M \oplus \Lambda^2_F
\end{equation}
with dimension
\begin{equation}
    \dim \Lambda^2 = \mathbf{3 \oplus 3}
\end{equation}
There is already a great deal that can be said about the structure of the Riemann matrix from this fact and the decomposition (\ref{Warp_Alg_Curv}). Considering an adapted orthonormal co-frame
\begin{equation}
    e^0 = dt, \quad e^i = f \eta^i, \quad i =1,2,3,
\end{equation}
(where $\eta^i$ is orthonormal on the fiber $F$), we take the natural basis in (\ref{1_plus_3_2_Forms_Split}) via the identification of the mixed and pure 2-forms
\begin{equation}\label{1_3_Basis}
    \begin{split}
        E_1 = e^0 \wedge e^1, \quad E_2 = e^0 \wedge e^2, \quad E_3 = e^0 \wedge e^3, \\
        F_1 = e^2 \wedge e^3, \quad F_2 = e^3 \wedge e^1, \quad F_3 = e^1 \wedge e^2,
    \end{split}
\end{equation}
Then, an arbitrary element $\mathcal{Q}$ of $S^2(\Lambda^2 T^* \mathcal{M}) \cong S^2(\Lambda^2_M \oplus \Lambda^2_F)$ is of the form
\begin{equation}\label{1_3_Block_Matrix}
    \mathcal{Q} = \begin{pmatrix}
        M & N \\
        N^T  & P
    \end{pmatrix},
\end{equation}
where 
\begin{equation}
    M \in S^2 \Lambda^2_M, \quad P \in S^2 \Lambda^2_F, \quad N \in \Lambda^2_M \otimes \Lambda^2_F.
\end{equation}
In total, the dimension of these spaces add to $21$ (as expected). The Riemann tensor, however, satisfies the first Bianchi identity which eliminates one of these. The constraint is on any element that can yield a 4-form. The only possible contribution comes from the off-diagonal block $N$, since
\begin{equation}
    E_i \wedge F_j = \delta_{ij} \; e^0 \wedge e^1 \wedge e^2 \wedge e^3.
\end{equation}
Thus, the Bianchi identity imposes 
\begin{equation}
    \text{tr} \; N = 0.
\end{equation}
We can then define the algebraic space of curvature tensors $\mathcal{K}$ as 
\begin{equation}
    \mathcal{K} = \bigg\{ \begin{pmatrix}
        M & N \\
        N^T & P
    \end{pmatrix}\colon M = M^T, P = P^T, \text{tr} \;N = 0\bigg\}.
\end{equation}

To determine the content of these blocks for the Riemann matrix $\text{Riem}_{ij}$, we require the actual structure of the warped product. The metric on $\mathcal{M}$ takes the form
\begin{equation}
    g = dt^2 + f(t)^2 g_F.
\end{equation}
Since $B$ is one-dimensional, its curvature (\ref{Warp_R_1}) vanishes. Similarly since $F$ is three-dimensional, its curvature  is completely determined by its Ricci tensor. The total curvature of $\mathcal{M}$ is then
\begin{gather}
        R_{0i}^{\quad 0j} = -\frac{f''}{f} \delta_{ij} \\
        R_{ij}^{\quad kl} = \frac{1}{f^2} R_{ij}^{F \;  kl} - \frac{2 (f')^2}{f^2}\delta_i^{[k} \delta^{l]}_j,
\end{gather}
where
\begin{equation}\label{3_1_Ric_Ref}
    R_{ij}^{F\; kl} = 2\big(\delta_{i}^{[k}R_{j}^{F\;l]} - \delta_{j}^{[k} R_{i}^{F \; l]} \big) - R_F\delta_{i}^{[k} \delta_{j}^{l]}
\end{equation}
From this, we see that the Riemann matrix structure is block diagonal,
\begin{equation}\label{Riem_Matrix_3_1}
    \text{Riem} = \begin{pmatrix}
        -\frac{f''}{f}\mathbb{I}_3 & 0\\
        0 & \mathcal{D}
    \end{pmatrix},
\end{equation}
where
\begin{equation}
    \mathcal{D} = \frac{1}{f^2} \mathcal{R}_F - \frac{(f')^2}{f^2}\mathbb{I}_3
\end{equation}
and $\mathcal{R}_F$ is the matrix representation of (\ref{3_1_Ric_Ref}), in analogy with (\ref{3d_Curv_Operators}).
Let us now consider the limit in which the warped product manifold is Einstein. We can understand this constraint by mapping to the chiral basis (\ref{2_Form_Decomp}) and subsequently demanding that the off-diagonal components of the resulting Riemann matrix ($\text{Riem}'$) vanish. In the basis (\ref{4d_Basis}), we note
\begin{equation}
    \star E_i = F_i, \quad \star F_i = E_i.
\end{equation}
Consequently, the chiral basis can be taken as 
\begin{equation}\label{1_3_Chiral}
    \Sigma^{\pm}_i = \frac{1}{\sqrt{2}} (E_i \pm F_i).
\end{equation}
The matrix that maps between these two is again (\ref{P_Matrix}), and under the similarity transformation
\begin{equation}
    \text{Riem}' = \mathcal{{P}} \; \text{Riem} \; \mathcal{P}^T,
\end{equation}
we obtain
\begin{equation}\label{Rie_Matrix_Chiral_1_3}
    \text{Riem}' = \frac{1}{2}\begin{pmatrix}
        -\frac{f''}{f} \mathbb{I}_3 + \mathcal{D} & -\frac{f''}{f} \mathbb{I}_3 - \mathcal{D} \\
        -\frac{f''}{f} \mathbb{I}_3 - \mathcal{D} & -\frac{f''}{f} \mathbb{I}_3 + \mathcal{D}
    \end{pmatrix}.
\end{equation}
For the off-diagonal blocks to vanish, we see that the fiber must itself be Einstein (\ref{Riem_3d_Einstein})
\begin{equation}\label{Einstein_Cond_1_3_Ref}
    \mathcal{R}_F = \alpha \mathbb{I}_3, \quad \alpha = (f')^2 -f f''
\end{equation}
and hence maximally symmetric\footnote{Furthermore for the fiber to be Ricci flat, the function $f$ must satisfy
\begin{equation}
    (f')^2 - f f'' = 0,
\end{equation}
yielding the de Sitter Universe:
\begin{equation}
    f(t) = c_1 e^{c_2 t},
\end{equation} which is internally consistent.}. Thus, a $1+3$ warped product is Einstein if the Riemann matrix is proportional to the identity
\begin{equation}\label{1_3_Rie_Fin}
    \text{Riem}_{ij} = -\frac{f''}{f} \mathbb{I}_{ij},
\end{equation}
where $f$ satisfies
\begin{equation}\label{1_3_Warp_Constraint_f}
    f'' + \frac{\Lambda}{3} f = 0.
\end{equation}
The reverse direction is also true: given a $1+3$ warped product with Riemann matrix of the form (\ref{1_3_Rie_Fin}), the manifold is Einstein: similarity transforming into the chiral basis (\ref{1_3_Chiral}), the matrix commutes with the Hodge dual and hence the manifold is Einstein.

\subsection{Case \RNum{2}: 2 + 2 Warping}
When both the base $B$ and the fiber $F$ of the warped product are two-dimensional, then the space of 2-forms $\Lambda^2$ on $\mathcal{M}$ contains the volume form on $B$, the volume form on $F$ and a four-dimensional subspace spanned by the mixed products of 1-forms on both the base and the fiber:
\begin{equation}
    \Lambda^2 \cong \Lambda^2_B \oplus \Lambda^2_F \oplus \Lambda^2_M.
\end{equation}
Again, prior to discussing the warp geometry the matrix structure of the curvature is constrained by the irreducible decomposition (\ref{Warp_Alg_Curv}). We consider the following basis on 2-forms,
\begin{gather}\label{2_2_Basis}
    \omega_B = e^1 \wedge e^2, \quad \omega_F = e^3 \wedge e^4, \\
    m_1 = e^1 \wedge e^3, \; m_2 = e^1 \wedge e^4, \; m_3 = e^2 \wedge e^3, \; m_4 = e^2 \wedge e^4.
\end{gather}
Labeling the elements of $S^2(\Lambda^2 T^*\mathcal{M}) \cong S^2(\Lambda^2_B \oplus \Lambda^2_F \oplus \Lambda^2_M)$ as 
\begin{equation}\label{2_2_Decomp}
    \underbrace{S^2(\Lambda^2_B)}_{\alpha} \oplus \underbrace{S^2 (\Lambda^2_F)}_{\beta} \oplus \underbrace{S^2(\Lambda^2_M)}_M\oplus
    \underbrace{(\Lambda^2_B \otimes \Lambda^2_F)}_{\gamma} \oplus \underbrace{(\Lambda^2_B \otimes \Lambda^2_M)}_u \oplus \underbrace{(\Lambda^2_M\otimes \Lambda^2_F)}_v
\end{equation}
with dimension
\begin{equation}
    \mathbf{1 \oplus 1 \oplus 10 \oplus 1 \oplus 4 \oplus 4}
\end{equation}
a general matrix $\mathcal{L} \in S^2(\Lambda^2 T^* \mathcal{M})$ takes the form
\begin{equation}
    \mathcal{L} = \begin{pmatrix}
        \alpha & \gamma & u^T \\
        \gamma & \beta & v^T \\
        u & v & M
    \end{pmatrix}.
\end{equation}
One now must impose the constraint from the first Bianchi identity on this matrix. The elements capable of inducing a top-form on $\mathcal{M}$ in (\ref{2_2_Decomp}) are $\gamma$ and the completely antisymmetric one-dimensional subspace within $M$ ($M_{14} - M_{23}$). In particular, the condition
\begin{equation}
    \gamma  - M_{14} + M_{23} = 0 
\end{equation}
defines the algebraic space of curvature tensors,
\begin{equation}
    \mathcal{K} \coloneqq \bigg\{ \begin{pmatrix}
        \alpha & \gamma & u^T \\
        \gamma & \beta & v^T \\
        u & v & M 
    \end{pmatrix} \bigg\vert  \; \gamma  - M_{14} + M_{23} = 0  \bigg\}.
\end{equation}

Since both $B$ and $F$ are two dimensional, their respective curvatures are completely determined by their Gaussian curvatures $K$. Due to the metric structure (\ref{Warp_Metric}), a number of the irreducible representations (\ref{2_2_Decomp}) immediately collapse,
\begin{gather}
    \gamma = u = v = 0, \\
    M_{12} = M_{14} = M_{23} = M_{43} = 0.
\end{gather}
The resulting matrix form is
\begin{equation}\label{2_2_Riem_Matrix}
    \text{Riem} = 
    \begin{pmatrix}
        K_B & 0 & 0 & 0 & 0 & 0 \\
        0 & \frac{K_F}{f^2} - \frac{\vert \nabla f\vert^2}{f^2} & 0 & 0 &0 & 0 \\
        0 & 0 & -\frac{f_{;11}}{f} & 0 & -\frac{f_{;12}}{f} & 0 \\
        0 & 0 & 0 & -\frac{f_{;11}}{f} & 0 & -\frac{f_{;12}}{f} \\ 0 & 0 & -\frac{f_{;12}}{f} & 0 & -\frac{f_{;22}}{f} & 0 \\
        0 & 0 & 0 & -\frac{f_{;12}}{f} & 0 & -\frac{f_{;22}}{f}
    \end{pmatrix}.
\end{equation}

In order to transform into the chiral decomposition (\ref{2_Form_Decomp}), we introduce the basis
\begin{equation}\label{2_2_Chiral}
    \Sigma^{\pm}_1 = \frac{\omega_B \pm \omega_F}{\sqrt{2}}, \quad \Sigma^{\pm}_2 = \frac{m_1 \pm m_4}{\sqrt{2}}, \quad \Sigma^{\pm}_3 = \frac{m_2 \pm m_3}{\sqrt{2}}
\end{equation}
with the matrix $\mathcal{P}$ mapping between (\ref{2_2_Basis}) and (\ref{2_2_Chiral}) of the form 
\begin{equation}
    \mathcal{P} = \frac{1}{\sqrt{2}} \begin{pmatrix}
        1 & 1 & 0 & 0 & 0 & 0 \\
        1 & -1 & 0 & 0 &0 & 0 \\
        0 & 0 & 1 & 0 & 0 & 1  \\
        0 & 0 & 0 & 1 & 1 & 0 \\
        0 & 0 & 0 & 1 & -1 & 0 \\
        0 & 0 & -1 & 0 & 0 & 1
    \end{pmatrix}
\end{equation}
Under a similarity transformation, the Riemann matrix (\ref{2_2_Riem_Matrix}) admits the block decomposition
\begin{equation}
    \text{Riem}' = \mathcal{P} \; \text{Riem} \; \mathcal
    P^T = \begin{pmatrix}
        \mathcal{E} & \mathcal{F} \\
        \mathcal{F}^T & \mathcal{E}
    \end{pmatrix}
\end{equation}
where
\begin{equation}\label{Diagonal_2_2}
    \mathcal{E} = \frac{1}{2}\begin{pmatrix}
        K_B + \frac{K_F}{f^2} - \frac{\vert \nabla f\vert^2}{f^2} & 0 & 0 \\
        0 & -\frac{f_{;11}}{f} -\frac{f_{;22}}{f} & 0 \\
        0 & 0 &  -\frac{f_{;11}}{f} -\frac{f_{;22}}{f}
    \end{pmatrix}
\end{equation}
and
\begin{equation}
    \mathcal{F} = \frac{1}{2} \begin{pmatrix}
        K_B - \frac{K_F}{f^2} + \frac{\vert \nabla f\vert^2}{f^2} & 0 & 0  \\
        0 & -\frac{f_{;11}}{f} +\frac{f_{;22}}{f} & \frac{2f_{;12}}{f} \\
        0 &  -\frac{2f_{;12}}{f} &  -\frac{f_{;11}}{f} +\frac{f_{;22}}{f}
    \end{pmatrix}.
\end{equation}
For the spacetime to be Einstein (\ref{Einstein_Cond}), the off-diagonal block $\mathcal{F}$ must vanish. This implies
\begin{equation}
    K_B = \frac{K_f- \vert \nabla f\vert^2}{f^2}, \quad f_{;11} = f_{;22}, \quad f_{;12} = 0.
\end{equation}
In words these conditions enforce the horizontal and vertical sectional curvatures of the full warped metric to agree, and the Hessian of the warping function to be pure trace on the base. In the warped product basis, the manifold is Einstein if the Riemann matrix takes the form
\begin{equation}\label{Riem_Matrix_Einstein_2_2} \text{Riem} = 
    \begin{pmatrix}
         K_B & 0 & 0 & 0 & 0 & 0 \\
        0 & K_B & 0 & 0 & 0 & 0 \\
        0 & 0 & -\frac{f_{;11}}{f} & 0 & 0 & 0 \\
        0 & 0 & 0 & -\frac{f_{;11}}{f} & 0 & 0 \\
        0 & 0 & 0 & 0 & -\frac{f_{;11}}{f} & 0 \\
        0 & 0 & 0 & 0 & 0 & -\frac{f_{;11}}{f}
    \end{pmatrix}.
\end{equation}
Again, the reverse direction is also true for the same reasons as the $1+3$ case.

\subsection{Case \RNum{3}: 3 + 1 Warp}
This case is an important comparison to case \RNum{1}, since although the structure of the 2-forms still has a $\mathbf{3 \oplus 3}$ decomposition (corresponding now to the 2-forms sourced from the three-dimensional base, and the mixed 2-forms from the 3 1-forms on $B$ and the single 1-form on the fiber $F$), the corresponding Einstein limits prove to be quite different. To the same end as (\ref{4d_Basis}), we introduce the adapted basis for $\Lambda^2$ as 
\begin{equation}\label{3_1_Basis}
    \begin{split}
        \mathcal{F}_1 = e^2 \wedge e^3, \quad \mathcal{F}_2 = e^3 \wedge e^1, \quad \mathcal{F}_3 = e^1 \wedge e^2, \\
        \mathcal{E}_1 = e^1 \wedge e^4, \quad \mathcal{E}_2 = e^2 \wedge e^4, \quad \mathcal{E}_3 = e^3 \wedge e^4,
    \end{split}
\end{equation}
where $e^4 = f dy$ is the fiber coordinate. The matrix representation of $S^2(\Lambda^2 T^*\mathcal{M})$ again has a block structure (\ref{1_3_Block_Matrix}), with the Bianchi identity requiring that the trace of the off-diagonal blocks vanishes. The warped product curvature receives the following contributions
\begin{equation}
    R_{abcd} = R^B_{abcd}, \quad R_{a4b4} = -\frac{f_{;ab}}{f}
\end{equation}
with all other components vanishing. Consequently in the basis ($\mathcal{F}_i, \mathcal{E}_i$), the Riemann matrix (\ref{Riemann_Map}) assumes the block diagonal form
\begin{equation}
    \text{Riem} = \begin{pmatrix}
        \mathcal{R}_B - \frac{R_B}{2}\mathbb{I}_3 & 0 \\
        0 & -\frac{\text{Hess}(f)}{f}
    \end{pmatrix},
\end{equation}
where $\mathcal{R}_B$ is the Ricci matrix on the base (\ref{3d_Curv_Operators}). Introducing the chiral basis as 
\begin{equation}\label{3_1_Chiral}
    \Sigma^{\pm}_i = \frac{1}{\sqrt{2}} (\mathcal{F}_i \pm \mathcal{E}_i),
\end{equation}
the transformation between the two bases is given simply by the matrix (\ref{P_Matrix}) which yields the block form
\begin{equation}
    \text{Riem}' = \mathcal{P} \; \text{Riem} \; \mathcal{P}^T = \frac{1}{2}\begin{pmatrix}
        \mathcal{R}_B - \frac{R_B}{2}\mathbb{I}_3 - \frac{\text{Hess}(f)}{f} & \mathcal{R}_B - \frac{R_B}{2}\mathbb{I}_3 + \frac{\text{Hess}(f)}{f} \\ \mathcal{R}_B - \frac{R_B}{2}\mathbb{I}_3 + \frac{\text{Hess}(f)}{f} & \mathcal{R}_B - \frac{R_B}{2}\mathbb{I}_3 - \frac{\text{Hess}(f)}{f}
    \end{pmatrix}
\end{equation}
Imposing that the off-diagonal blocks vanish enforces the Einstein condition. This amounts to demanding that the curvature of the base be (minus) the Hessian acting on the mixed forms $e^{a4}$. For an Einstein $3+1$-warped product, in both the chiral basis (\ref{3_1_Chiral}) and the adapted warped product basis (\ref{3_1_Basis}) the Riemann matrix has the block diagonal form
\begin{equation}
    \text{Riem} = \begin{pmatrix}
        \mathcal{R}_B - \frac{R_B}{2}\mathbb{I}_3 & 0 \\
        0 & \mathcal{R}_B - \frac{R_B}{2}\mathbb{I}_3
    \end{pmatrix}
\end{equation}
Compare this to the $1+3$ case (\ref{1_3_Rie_Fin}), in which the Riemann matrix is diagonal $\text{Riem}_{ij} = -f''/f \; \mathbb{I}_{ij}$ in the Einstein limit.

\section{Immediate Consequences}
The ability to similarity transform the Riemann matrix between alternative representations of the space of algebraic curvature tensors in four dimensions is a powerful tool. Since a number of results are known from the chiral decomposition of the 2-forms (\ref{2_Form_Decomp}) - typically of much simpler form than the equivalent statements adapted to the metric \cite{atiyah_self-duality_1978, krasnov_formulations_2020, shaw_general_2026} -  these can be translated into statements regarding the geometry of the warp in a simple way. One can, for instance, explore the Petrov classification via isolating the self-dual component of the Weyl tensor \cite{petrov_classification_2000, stephani_exact_2003, kroon_conformal_2023}. One sees in this way that the four dimensional Einstein warped products are naturally ordered according to their algebraic specialization.

\subsection{Petrov Classification of Four-dimensional Warped Products}
Under the decomposition (\ref{2_Form_Decomp}), the Weyl tensor $C_{\mu \nu \rho \sigma}$ splits \cite{besse_einstein_1987, hughes_encoding_2026} into self-dual and anti-self-dual elements (\ref{Weyl_Matrix})
\begin{equation}
    C = C^+ \oplus C^-.
\end{equation}
The Petrov classification \cite{petrov_classification_2000, felice_relativity_1990} classifies the algebraic type of the operator 
\begin{equation}
    C^+\colon \Lambda^+ \to \Lambda^+
\end{equation}
(or, equivalently, the totally symmetric Weyl spinor $\Psi_{ABCD}$ \cite{stewart_advanced_1991}). The similarity transformations on the space $\mathcal{K}$ provide an efficient means of constructing this operator and consequently the available algebraic types.

In the $2+2$ case, the diagonal block of the Riemann tensor in the chiral basis (\ref{2_2_Chiral}) was given in (\ref{Diagonal_2_2}) as 
\begin{equation}
    \mathcal{E} = \frac{1}{2}\begin{pmatrix}
        K_B + \frac{K_F}{f^2} - \frac{\vert \nabla f\vert^2}{f^2} & 0 & 0 \\
        0 & -\frac{f_{;11}}{f} -\frac{f_{;22}}{f} & 0 \\
        0 & 0 &  -\frac{f_{;11}}{f} -\frac{f_{;22}}{f}
    \end{pmatrix}.
\end{equation}
In general (\ref{Chiral_Matrix}), this matrix contains contributions from both $W^+$ and the scalar curvature. Consequentially, we can extract $W^+$ via the identification
\begin{equation}
    C^+ = \mathcal{E} - \frac{\text{tr} \; \mathcal{E}}{3} \mathbb{I}_3.
\end{equation}
Introducing
\begin{equation}
    a = \frac{1}{2}\bigg(K_B + \frac{K_F}{f^2} - \frac{\vert \nabla f\vert^2}{f^2}\bigg), \quad b = \frac{1}{2} \bigg(-\frac{f_{;11}}{f} -\frac{f_{;22}}{f}\bigg)
\end{equation}
we see
\begin{equation}
    C^+ = \text{diag} \bigg(\frac{2}{3} (a-b), -\frac{1}{3} (a-b), -\frac{1}{3}(a-b)\bigg).
\end{equation}
The eigenvalue pattern of the self-dual Weyl tensor is thus
\begin{equation}
    (\lambda, -\lambda/2, -\lambda/2)
\end{equation}
which is precisely Petrov type-D. This can degenerate to type-O in the limit $\lambda\to 0$, which is distinct from the Einstein condition (\ref{Riem_Matrix_Einstein_2_2}).

In the $1+3$ case, the algebraic structure is richer since the trace-free Ricci tensor of the fiber sources the self-dual Weyl tensor. Indeed, from (\ref{Rie_Matrix_Chiral_1_3}) we see that 
\begin{equation}
    C^+ = \frac{1}{2f^2} \bigg(\mathcal{R}_F - \frac{\text{tr}\; \mathcal{R}_F}{3} \mathbb{I}_3\bigg).
\end{equation}
As such, the Petrov type is determined by the eigenvalue structure of the fiber itself:
\begin{enumerate}
    \item If the trace-free Ricci tensor $Z_F$ of $F$ has three distinct eigenvalues, then the warped product spacetime is type-\RNum{1}. This cannot degenerate to type-\RNum{2} since as a real symmetric endomorphism on $T_pF$, $Z_F$ is diagonalizable. 
    \item If $Z_F$ has one repeated eigenvalue, then the manifold is type-D.
    \item For the manifold to be type-O, $Z_F$ must vanish meaning that the fiber is Einstein (and hence a space form). However, we previously established this to be the necessary condition for the manifold itself to be Einstein (\ref{Einstein_Cond_1_3_Ref}). Thus, $1+3$ warped products are of type-O if and only if they are Einstein. 
\end{enumerate}
Contrast this with the $3+1$ case, in which $C^+$ is 
\begin{equation}
    C^+ = \frac{1}{2} \bigg[ \bigg(\mathcal{R}_B - \frac{R_B}{2}\mathbb{I}_3 - \frac{\text{Hess}(f)}{f}\bigg) - \frac{1}{3}\text{tr}\bigg(\mathcal{R}_B - \frac{R_B}{2}\mathbb{I}_3 - \frac{\text{Hess}(f)}{f}\bigg) \mathbb{I}_3 \bigg].
\end{equation}
While generically this can be type-\RNum{1}, D or O (depending on the eigenvalue structure, as before), note that in the Einstein limit \textit{this remains true}:
\begin{equation}
    C^+ =  \bigg(\mathcal{R}_B - \frac{R_B}{2}\mathbb{I}_3\bigg) - \frac{1}{3}\text{tr}\bigg(\mathcal{R}_B - \frac{R_B}{2}\mathbb{I}_3 \bigg) \mathbb{I}_3.
\end{equation}
This allows us to establish a algebraic hierarchy for four-dimensional warped product Einstein manifolds: $3+1$ are generically type-\RNum{1}, $2 + 2$ are generically type-D and $1 + 3$ are type-O (note that - up to the discussion of Einstein manifolds - these results recover those reported previously by Carot and de Costa using holonomy methods \cite{carot_geometry_1993}).

\subsection{Topological Constraints in the Compact Case}
The constraint $C^+ = 0$ (often referred to as half-conformally flat) is particularly important in four-dimensional geometry, as it corresponds to the limit in which the twistor structure of Penrose is integrable \cite{besse_einstein_1987, mason_integrability_1996, eguchi_gravitation_1980}. These are the type-O Petrov classes. For each of the cases we have considered above (regardless of metric signature) $C^+ = C^-$ since the diagonal blocks of the Riemann matrix in the chiral basis are always the same. This is perhaps the strongest constraint on a warped product manifold in four dimensions, since demanding the manifold be half-conformally flat amounts to it being fully conformally flat. If one additionally imposes the Einstein condition, then from the orthogonal decomposition (\ref{Irreducible_Decomp_Riem}) $\mathcal{M}$ must be constant curvature. Actually the $1+3$ warp is special in this regard, since as we have seen enforcing Einstein alone is enough to yield constant curvature.

One can also say something regarding the topological structure of Einstein warped products. In what follows let $\mathcal{M}$ be a closed (compact without boundary) Riemannian manifold. Then the first Pontryagin class $p_1$ (via the Chern-Weil theorem) and the Euler characteristic $\chi$ (via the Chern-Gauss-Bonnet theorem) can be expressed as integrals over certain matrix elements in the chiral decomposition (\ref{Chiral_Matrix}) \cite{besse_einstein_1987, chern_curvatura_1945}:
\begin{equation}\label{CBG}
    \begin{split}
        p_1(M) & = \frac{1}{4\pi^2} \int_{\mathcal{M}} \big(\vert C^+ \vert^2 - \vert C^- \vert^2\big) \text{vol}_g, \\
        \chi(M) & = \frac{1}{8\pi^2} \int_{\mathcal{M}} \bigg(\vert C^+ \vert^2 + \vert C^- \vert^2 + \frac{R^2}{24}\bigg) \text{vol}_g
    \end{split}
\end{equation}
Consequently, we have
\begin{equation}
    2\chi \geq \vert p_1\vert 
\end{equation}
which results more generally from combining the Hirzebruch signature theorem \cite{hirzebruch_topological_1966} with the Thorpe-Hitchin inequality \cite{thorpe_remarks_1969, hitchin_compact_1974}. For the four-dimensional Einstein warped products, one has (for all cases)
\begin{equation}\label{Euler_Einstein_Warp}
    \chi(\mathcal{M}) = \frac{1}{8\pi^2}\int_{\mathcal{M}} \bigg(2\vert C^+ \vert^2  + \frac{R^2}{24}\bigg) \text{vol}_g  \geq 0
\end{equation}
since $p_1 = 0$. However, since $\mathcal{M}$ is fundamentally a product manifold $\mathcal{M} = B \times_f F$ (where $f$ is a smooth positive function), we also have that \cite{eguchi_gravitation_1980, nakahara_geometry_2003} 
\begin{equation}
    \chi(\mathcal{M}) = \chi(B) \chi(F).
\end{equation}
Some interesting consequences follow from considering each of the four-dimensional Einstein warped cases in turn.
\begin{enumerate}
    \item \textit{Case \RNum{1}}: When $\mathcal{M}$ is $1+3$ the Euler characteristic vanishes, $\chi(F) = \chi(B) = 0$ (since $\mathcal{M}$ is closed and $F$ and $B$ are odd-dimensional). However, since in this case Einstein enforces conformal flatness (\ref{Euler_Einstein_Warp}) gives
    \begin{equation}\label{1_3_Euler}
        \frac{R^2}{192 \pi^2} \text{vol}(\mathcal{M}) = 0
    \end{equation}
    and hence the manifold is flat (one example here being the 4-torus). This can also be seen from the Einstein condition on the warping function (\ref{1_3_Warp_Constraint_f}), 
    \begin{equation}
        f'' + \frac{\Lambda}{3}f = 0.
    \end{equation}
    For the base to be closed, it is essentially $S^1$. The warping function must be smooth, positive \textit{and} periodic. The only possibility is that $\Lambda = 0$ and $f$ is a constant, which is precisely what the global argument (\ref{1_3_Euler}) demands at the characteristic level.
    \item \textit{Case \RNum{2}}: When $\mathcal{M}$ is $2+2$, its Euler characteristic is determined by the genus $g$ of the base and the genus $h$ of the fiber via
    \begin{equation}
        (2-2g)(2-2h) = \frac{1}{8\pi^2} \int_{\mathcal{M}} \bigg[\frac{16}{3}\bigg(K_B - \frac{f_{;11}}{f}\bigg)^2  + \frac{2}{3} \bigg(K_B + 2\frac{f_{;11}}{f} \bigg)^2\bigg] \text{vol} \geq 0.
    \end{equation}
    Suppose that $\mathcal{M}$ is also half-conformally flat (hence conformally flat). One has (since $R = 4(K_B + 2f_{;11}/f)$)
    \begin{equation}
        (2-2g)(2-2h) = \frac{R^2}{192 \pi^2} \text{vol} (\mathcal{M}) \geq 0.
    \end{equation}
    while in contrast to the $1+3$ case it appears that $2+2$ Einstein can support positive, zero and negative scalar curvature in the (half-) conformally flat limit, it is in fact the case that only zero $(T^2 \times T^2)$ survives. The reason that $R>0$ is excluded is a consequence of Hitchin's theorem \cite{besse_einstein_1987, hitchin_compact_1974}: a compact half-conformally flat Einstein manifold in four dimensions with positive scalar curvature must be isometric to either the 4-sphere $S^4$ or $\mathbb{C}P^2$ with their canonical metrics. Neither of these are a possibility under the above constraints. Negative $R$ (namely, a hyperbolic manifold) is excluded because the problem reduces to finding constant negative curvature metrics that are the products of regular surfaces, which is impossible (a four-dimensional hyperbolic manifold cannot have fundamental group splitting as $\pi_1(\Sigma_g) \times \pi_1(\Sigma_h)$, whereas a regular product of surfaces does \cite{bridson_metric_1999, hatcher_algebraic_2002}). This leaves only the flat case, which under the constraint $\chi(B)\chi(F) = 0$ leaves essentially $T^2 \times T^2$.
    \item \textit{Case \RNum{3}}: If $\mathcal{M}$ is $3+1$, then the Euler characteristic again vanishes and the (half-) conformally flat limit reduces to $R = 0$ for the same reasons as in case \RNum{1}, leaving in the simplest case $T^4$.
\end{enumerate}

\subsection{A Comment on Plebanski}
It is worth emphasizing that the approach laid out in section 3 is equivalent to the solving of the Plebanski equations (in the absence of energy-momentum) for the warp geometry \cite{plebanski_separation_1977, krasnov_plebanski_2011}. In Plebanski's formulation, one uses the fact that the 2-forms decompose into the eigenspaces of the Hodge dual (\ref{2_Form_Decomp}) and thus the curvature (as a 2-form) also decomposes as (\ref{Chiral_Matrix})
\begin{equation}
    F^i = \mathcal{A}^i_j \Sigma^{+}_j + \mathcal{B}^i_j \Sigma^{-}_j.
\end{equation}
The Plebanski field equation (which can be recovered from a variational principle \cite{krasnov_formulations_2020, capovilla_self-dual_1991}) is the statement that the curvature $F^i$ is purely self-dual as a 2-form, which is equivalent to the vacuum Einstein equations or the statement that the manifold is Einstein (\ref{Einstein_Cond}). In other words, $\mathcal{B}^i_j = 0$ and the Riemann tensor (\ref{Chiral_Matrix}) must be block diagonal. When we construct the Riemann operator in the warped product basis and similarity transform to the chiral basis, this is the condition we enforce. The ability to do this is unique to four dimensions, and we have seen how exploiting this fact makes the Einstein limit of warped products particularly transparent.

\section{Conclusions}
By viewing the Riemann tensor as an endomorphism operator on the space of 2-forms, one gets in four dimensions a $6\times 6$ matrix representation of curvature on the algebraic space of curvature tensors (\ref{Irreducible_Decomp_Riem}). By constraining this matrix relative to different bases on the 2-forms, one can impose various geometrical constraints on the manifold. Working with a basis adapted to the eigenspaces of the Hodge dual (\ref{Chiral_Matrix}), the Einstein limit is equivalent to this matrix being block diagonal \cite{besse_einstein_1987, krasnov_formulations_2020, hughes_encoding_2026}. The Plebanski reformulation of GR imposes this condition variationally via a modified BF theory \cite{krasnov_formulations_2020, capovilla_self-dual_1991, shaw_general_2026}. When working with warped products, this fact offers an economical approach to determining the Einstein limit in each of the available cases via similarity transformations to the chiral basis. We summarize this into the following the theorem:

\begin{theorem}
Let $\mathcal{M}$ be a four-dimensional warped product (pseudo-) Riemannian manifold, with warping function $f$. In the $1+3$ case, the manifold is Einstein if and only if the Riemann matrix is diagonal and of the form,
\begin{equation}
    \mathrm{Riem}_{ij} = -\frac{f''}{f} \mathbb{I}_{ij}.
\end{equation}
In the $2+2$ case, the manifold is Einstein if and only if the Riemann matrix is diagonal and of the form  \begin{equation} \mathrm{Riem} = 
    \begin{pmatrix}
         K_B & 0 & 0 & 0 & 0 & 0 \\
        0 & K_B & 0 & 0 & 0 & 0 \\
        0 & 0 & -\frac{f_{;11}}{f} & 0 & 0 & 0 \\
        0 & 0 & 0 & -\frac{f_{;11}}{f} & 0 & 0 \\
        0 & 0 & 0 & 0 & -\frac{f_{;11}}{f} & 0 \\
        0 & 0 & 0 & 0 & 0 & -\frac{f_{;11}}{f}
    \end{pmatrix}.
\end{equation}
In the $3+1$ case, the manifold is Einstein if and only if the Riemann matrix is block diagonal and of the form
\begin{equation}
    \mathrm{Riem} = \begin{pmatrix}
        \mathcal{R}_B - \frac{R_B}{2}\mathbb{I}_3 & 0 \\
        0 & \mathcal{R}_B - \frac{R_B}{2}\mathbb{I}_3
    \end{pmatrix}.
\end{equation}
\end{theorem}
Since a number of results in four dimensional geometry leverage the chiral structure of the 2-forms (via, for instance, the properties of the self-dual Weyl tensor) the observations above lend themselves naturally to further classifications of warped products. This leads to the following:
\begin{theorem}
    Let $\mathcal{M}$ be a four-dimensional Einstein warped product (pseudo-) Riemannian manifold. Then $3+1$ warps are generically of Petrov type-\RNum{1}, $2+2$ are generically of Petrov type-D, and $1+3$ are conformally flat and of Petrov type-O.
\end{theorem}

Furthermore, in the compact Riemannian case the same algebraic structure has immediate topological consequences. Since the warped-product metrics considered here are regular products at the level of the underlying manifold, the Euler characteristic factorizes as $\chi(\mathcal{M}) = \chi(B) \chi(F)$. Combining this with the chiral Chern-Gauss-Bonnet formula (\ref{CBG}) shows that half-conformally flat Einstein limit is severely constrained. In the $(1+3)$ and $(3+1)$ cases the odd-dimensional factor forces $\chi(\mathcal{M}) = 0$, and these manifolds are consequently flat. In the $2+2$ case, the Euler characteristic is instead $(2-2g)(2-2h)$, so the topology appears immediately richer, but the half-conformally flat Einstein limit again reduces to the flat torus product $T^2 \times T^2$ under the closed regular product assumptions. Thus the three warped-product splittings differ algebraically in their Einstein and Petrov structure, but their compact half-conformally flat limits collapse to flat geometries. These results are summarized in Table (\ref{tab:warped-product-summary}).

\begin{table}[t]
\centering
\renewcommand{\arraystretch}{1.35}
\resizebox{\textwidth}{!}{%
\begin{tabular}{|c|c|c|c|}
\hline
\textbf{Warp} 
& \textbf{Einstein Riemann matrix} 
& \textbf{Einstein Petrov type} 
& \textbf{Closed HCF limit} \\
\hline
$1+3$ 
& $-\dfrac{f''}{f} I_6$ 
& $O$ 
& flat, e.g. $T^4$ \\
\hline
$2+2$ 
& $\operatorname{diag}\!\left(
K_B,K_B,
-\dfrac{f_{;11}}{f},
-\dfrac{f_{;11}}{f},
-\dfrac{f_{;11}}{f},
-\dfrac{f_{;11}}{f}
\right)$ 
& $D$  
& $T^2\times T^2$ \\
\hline
$3+1$ 
& $\begin{pmatrix}
Q & 0\\
0 & Q
\end{pmatrix}$,
\quad
$Q=\mathcal{R}_B-\dfrac{R_B}{2}I_3$
& $I$ 
& flat, e.g. $T^4$ \\
\hline
\end{tabular}%
}
\caption{Summary of the Einstein limits of four-dimensional warped products. Here HCF denotes the half-conformally flat limit. In the $3+1$ case, $Q$ is the common $3\times3$ curvature block appearing in the warped-product basis. In the closed Riemannian regular product setting, the HCF Einstein limits reduce to flat geometries.}
\label{tab:warped-product-summary}
\end{table}

\section{Acknowledgements}
Jack C. M. Hughes is grateful to Prof. Glenn Muschert for his support.

\section{Funding}
This work received institutional funding from Khalifa University. Jack C. M. Hughes and Fedor V. Kusmartsev acknowledge support from the Khalifa University Research and Innovation Grant KU-INT-RIG-2024-046/8474000759. Joudy F. Jamal Beek and Fedor V. Kusmartsev acknowledge support from the Khalifa University Research and Innovation Grants KU-INT-RIG-2023-8474000546 and KU-INT-RIG-2024-8474000754. Fedor V. Kusmartsev also acknowledges support from the Thousand Talents Program and the President’s International Fellowship Initiative of the Chinese Academy of Sciences Awards.

\section{Conflict of Interest}
The authors of this work declare that they have no conflicts of interest.

\section{References}
\bibliographystyle{iopart-num}
\bibliography{references.bib}

\end{document}